# Destruction of multiferroicity in $Tb_2BaNiO_5$ by Sr-doping and its implication to magnetodielectric coupling

**Sanjay K. Upadhyay and E.V. Sampathkumaran**

Tata Institute of Fundamental Research, Homi Bhabha Road, Colaba, Mumbai 400005, India

**Abstract:**
The Haldane spin-chain compound, $Tb_2BaNiO_5$, with two antiferromagnetic transitions, one at $T_1$= 63 K and the other at $T_2$= 25 K, has been recently shown to be an exotic multiferroic below $T_2$. Here, we report the results of our investigation of Sr doping at the Ba site by magnetization, heat-capacity, magnetodielectric (MDE) and pyrocurrent measurements. An intriguing finding, which we stress, is that the ferroelectricity is lost even for a doping level of 10 atomic percent, though magnetic ordering prevails. The doped specimens however retain significant magnetodielectric behaviour, but with reduced magnitudes and qualitative changes with respect to the behaviour of the parent compound. This implies that ferroelectric order is also crucial for the anomalously large MDE in the parent compound, in addition to the role of 4f single-ion anisotropy.

The topic of ferroelectricity due to certain types of magnetic structure are currently an important area in condensed matter physics. The materials which show such a behavior are named as 'type-II multiferroics' in the literature [1]. $TbMnO_3$ was the first example in this category, with two antiferromagnetic transitions, one at 41 K and the other at 28 K, and the lower antiferromagnetic transition triggers the ferroelectric transition [2]. Spin-driven multiferroic behavior has been reported not only in such 3-dimensional (3D) systems, but also in 2D (e.g., $Ni_2V_2O_8$, $CuFeO_2$) and 1D (e.g., $Ca_3CoMnO_6$, $LiVCuO_4$ etc.) magnetic systems [3 - 5]. Considering that quasi low dimensional systems are of great current interest to understand quantum phenomena (superconductivity, quantum spin liquid states etc.) at low temperatures [6], we have been probing multiferroicity and magnetodielectric (MDE) behavior of the spin-chain oxides in recent years.

In this respect, we placed particular emphasis on the prototype Haldane spin-chain [7] family, $R_2BaNiO_5$ (R= Gd, Dy, Er, Sm, Nd, Ho and Tb) [8 - 11], crystallizing in a centrosymmetric orthorhombic structure (*Immm* space-group). In this family, spin-chains are made up of $Ni^{2+}$ running along a-axis [12, 13]. It turned out that the behavior of the Tb compound in this family is quite exceptional. This compound exhibits the largest value of MDE coupling (18%) within this series following metamagnetic transition, which is rare among polycrystalline multiferroic compounds [11]. In this material, we found two antiferromagnetic transitions (one at $T_1$= 63 and the other at $T_2$= 25 K), with the magnetic transition at the lower temperature (25 K) inducing ferroelectricity. Though there is a collinearity of magnetic moments of Ni and Tb within respective sublattices, 3d(Ni) and 4f(Tb) moments were known to be mutually canted [14]. The readers may see Supplementary Material of Ref. 15 for all the parameters pertinent to crystal structure and magnetic structure of this compound, following Rietveld refinement of the neutron diffraction data. An intriguing finding we reported [15] is that this canting angle undergoes a sharp change at $T_2$ with exchangestriction anomalies, thereby pointing to the need for a new theory invoking critical canting angle, also involving 4f-orbital, for multiferroicity. It however appears that Haldane gap is not a necessary condition for the multiferroic anomalies, considering that the isostructural $Tb_2BaCoO_5$ is also multiferroic with still enhanced MDE coupling [16]. In order to enable better understanding of the novel coupling between magnetic and electric dipoles of the Ni compound, it is important to carry out studies on materials doped at different sites. Our recent investigations [17] on the series, $Tb_{2-x}Y_xBaNiO_5$, reveal that multiferroicity is retained even at the Y-end (x= 1.5), with respective transition temperatures varying essentially linearly with x (also see, Fig. S3 of Supplementary File in Ref. 15) and with profound MDE anomalies. This finding emphasized the involvement of Tb in electric dipole order [17], unlike in other rare-earth-based multiferroics known in the literature. The Y-doping studies also brought out the role of local (due to Tb 4f anisotropy) effects on MDE coupling. The aim of the present investigation is to dope at the Ba site (by Sr), that is, on the solid solution, $Tb_2Ba_{1-x}Sr_xNiO_5$, to see how multiferroicity is modified. This substitution is isovalent, but there is around 14% mismatch in ionic radii. [For $Ba^{2+}$ and $Sr^{2+}$, ionic radii are 1.35 Å and 1.18 Å respectively]. We have carried out magnetic, dielectric, ferroelectric and MDE measurements on the polycrystalline form of these compositions. The most intriguing finding is that Sr substitution for Ba, even for a small doping level, destroys ferroelectricity, in contrast to the observation on Y-based solid solution, however with a persistence of MDE with a reduced magnitude. This result brings out the role ferroelectric order on MDE, apart from local effects.

Polycrystalline $Tb_2Ba_{1-x}Sr_xNiO_5$ (x= 0.1, 0.2, 0.3, 0.4 and 0.5) specimens were prepared by a standard solid-state reaction method. Stoichiometric amounts of $Tb_2(CO_3)_2.nH_2O$ (99.9%), NiO (99.995 %), $BaCO_3$ (99.997%) and $SrCO_3$ (99.9% ) had been used as initial precursors. All precursors before mixing together were pre-sintered at 150 $^0$C for 2 hours to remove moisture, if any. The mixture was first calcined at 1200 $^0$C for 20 hours, followed by sintering at 1300 $^0$C for 30 hours after pelletizing, with intermediate grindings. The final sintering was done at 1350 $^0$C for 20 hours. All the heat treatments were carried out in Ar flow, because of the instability of $Tb^{3+}$ at high temperatures, according to Ref. 12. X-ray diffraction (XRD) (Cu-$K_\alpha$) measurements were performed to characterize the samples. Dc magnetic susceptibility ($\chi$) and isothermal magnetization (*M*) measurements were carried out by a commercial superconducting quantum interface device (SQUID, Quantum design, USA). Heat capacity (*C*) measurements were done with the help of a commercial Physical Properties Measurement System (PPMS, Quantum Design, USA). Agilent E4980A LCR meter with a home-made sample holder [8] integrated with the PPMS was used for the complex dielectric permittivity measurements with different frequencies ($\nu$ in the range 5 to 100 kHz). Conventional pyrocurrent ($I_{Pyro}$) and bias pyrocurrent ($I_{Bias}$) measurements were done with the help of a Keithley 6517B electrometer with the same sample holder as mentioned above. Isothermal dielectric constant ($\varepsilon'$) as a function of magnetic field (*H*) was measured at some selected temperatures (*T*).

In the x-ray powder diffraction (XRD) patterns, the observed diffraction lines could be indexed to orthorhombic structure (in *Immm* space group) up to x= 0.4 as shown in figure 1. As shown in figure 1e, for x= 0.5, additional lines start appearing and therefore this composition was not taken up for detailed studies. The XRD patterns of single phase compositions were subjected to Rietveld refinement by using Fullprof program [18]. As expected, there is a monotonous decrease in the unit-cell parameters, *a*, *b*, *c* and unit-cell volume (*V*) with respect to that of the parent compound as shown in figure 1(f). Despite that the solubility, as inferred from XRD, apparently extends up to x= 0.4, we restrict our discussions up to x= 0.2 only. The reason is that other apparent single phase compositions (x= 0.3 and 0.4) tend to show features in the measured properties attributable to the parent compound, superimposed over that of x= 0.2, as though there could be some phase segregation beyond x= 0.2 undetectable by XRD. But the present XRD patterns could not resolve the coexistence of these two compositions, possibly due to closeness of the lattice constants.

Figure 2(a) shows $\chi$(T) data below 150 K, obtained with *H*= 5 kOe, for the parent compound (taken from Ref. 11) and the doped compositions. The feature due to the onset of magnetic ordering at $T_1$ (shown by a dotted line in the figure 2a) is found to get suppressed to a lower temperature (near 55 K and 50 K for x= 0.1 and 0.2 respectively) for the doped compositions. The broad hump, usually attributed to 1D Haldane behavior, is retained. However, the feature at $T_2$ is not clearly resolvable in the $\chi$(T) data, but can be found in the derivative curves (figure 2b). As in the parent compound, a linear regime above 75 K is observed in the plot of inverse $\chi$(*T*) (not shown) and a fitting to Curie-Weiss law yields a value of 9.84 $\mu_B$/formula-unit (x= 0.1) and 9.70 $\mu_B$/formula-unit (x= 0.2) to the effective magnetic moment, which is close to that of free $Tb^{3+}$ ion; the value of the Curie paramagnetic temperature ($\theta_p$) turns out to be -19 K (for both), with the negative sign suggesting dominant antiferromagnetic correlations for the studied compositions. We have also recorded low-field (100 Oe) $\chi$(T) curves for the zero-field-cooled (ZFC) and field-cooled (FC) conditions and the curves are found to overlap (insets of figure 2(c) and (d)). This rules out the possibility of spin-glass behavior due to doping. In figures 2c and 2d, we show the isothermal magnetization behavior for doped compositions at two temperatures, 5 and 30 K. It is found that the metamagnetic transition field ($H_c$) is marginally suppressed due to doping, while retaining overall features of the parent compound (inset of Fig. 2d). The value of $H_c$ (defined as the field at which there is a sudden change in the slope of the curve) is 59.3 kOe and 59 kOe for x= 0.1 and 0.2 compositions respectively, while the value for the parent compound is $H_c$= 61.2 kOe

Further, in order to confirm the influence of Sr substitution on the magnetic transitions, we did heat capacity measurements. Figure 3(a) shows *C*(*T*)/*T* data for both the compositions along with that for the parent compound. From this data, one can clearly see that there is a qualitative change in the features due to doping. For x= 0.1, there is an anomaly near $T_1$ as in the parent compound, but the peak at 25 K is washed out; instead a sudden increase in slope is seen around ($T_2$=) 14 K, as a signature of another weak magnetic transition. For x= 0.2, the features due to both the transitions get more smeared.

We performed detailed dielectric measurements in various protocols [as a function of frequency ($\nu$), temperature and magnetic field]. Figs. 4(a) and (b) show the dielectric data as a function of temperature measured in the frequency range 5-100 kHz. Dielectric data shows a broad peak around 14 K and 6 K for the x= 0.1 and x= 0.2 compositions respectively. However, there is a weak frequency dispersion below and above the peak; this is in contrast to that observed in the parent compound, in which a $\nu$-independent $\lambda$-type peak was reported by us at a higher temperature ($T_2$ =25 K) as shown in the inset of fig. 4b. Possibly Sr doping tends to induce a marginal relaxor-type behavior, due to generation of polar nano-regions after Sr substitution. Similar generation of the polar nano region has also been reported for the 10% Al substituted (at the Mn site) of $TbMnO_3$ [19]. Finally, a careful inspection of Fig. 4a reveals that the peak-temperature shifts towards a lower temperature region with increasing $\nu$ in contrast to the expectation that it should increase with temperature; for 100 kHz curve, the peak temperature is lowered by about 1 K when compared to that for 5 kHz. The root cause of this discrepancy is not clear to us at present. Possibly, different nano regions depending on local chemical environment respond differently to frequency change in this particular material.

To see the MDE [defined as $\{\varepsilon'(H)-\varepsilon'(0)\}/\varepsilon'(0)$] behavior, $\varepsilon'(T)$ in the presence of different magnetic fields ($H$= 0, 40, 80 and 120 kOe) was measured. The results are shown in the mainframe of the figures 4(c) and 4(d) for the two Sr-doped compositions. For the Sr-doped compositions, it is interesting to note that, in $H$= 40 kOe (that is, below metamagnetic transition field), $\varepsilon'$ decreases distinctly at temperatures below that of the dielectric peak, establishing that the sign of MDE is negative. For the parent, however, the sign remains positive. For higher fields, say, 80 kOe, the sign of MDE becomes positive and the peak in $\varepsilon'(T)$ shifts to a lower temperature. To verify the above mentioned MDE behavior, we performed isothermal measurements of $\varepsilon'(H)$ up to 140 kOe at selected temperature (5, 10, 20 and 45 K). $\varepsilon'(H)$ for x= 0.1, though relatively small in magnitude, is distinctly negative in sign up to the field at which metamagnetic transition sets in (see the inset of figure 5a) at 5 and 10 K. Such a behavior is absent above $T_2$, e.g., for 20 K in this lower magnetic field regime, though, at higher fields, say, around 125 kOe, it gets negative again. There is also a peak in the plot of MDE at higher fields (see figure 5a) below $T_2$ and this peak gets marginally shifted to lower fields with increasing temperature. There is an additional sign reversal (to negative zone) at much higher fields (beyond 120 kOe). The above features are also observed for the x= 0.2 composition as shown in the figures 4(d) and 5(b). It may be noted that MDE remains positive for the parent compound in the measured range of magnetic field at all temperatures below $T_2$; the curve at 5 K reported in Ref. 11 is reproduced in the inset of Fig. 5b to demonstrate this. These results bring out the existence of various competing contributions (between possible antisymmetric part at low fields and symmetric part at high fields) to MDE due to Sr doping, while establishing that all these compositions are magnetodielectric. Complex MDE behavior with sign reversals are seen even in Y-doped specimens [17]. Finally, it is worth noting that the largest magnitude of MDE for the doped specimen is about 6.5% and 2.5 % (at 5 K in about 100 kOe, see the figures 5a and 5b). These values are large among polycrystalline form of materials. However, these are much smaller than that noted for the parent, which implies that the factors determining multiferroic order could play a role in determining unusually large MDE in the parent compound. Additionally, single-ion effects of Tb play a role, as emphasized earlier on the parent [11] and Y-doped systems [17]. Finally, the curves are hysteretic below $T_2$, as in the case of $M(H)$, supporting the idea of a coupling between magnetic and electric dipoles.

To search for ferroelectric behavior, we did detailed pyroelectric measurements with various protocols (conventional, bias electric method and rate dependence etc.). Figs. 6(a) and (b) show the conventional $I_{Pyro}$ data for the doped compositions. From the plots, one can see that, say, for the heating rate ($dT/dt$) of 2 K/min, for x= 0.1, there is a peak around 53.5 K when the sample is poled with an electric-field ($E$=) of -3.5 kV/cm. This peak can be reversed if the polarity of the applied electric field is reversed, as shown in upper inset of Fig. 6(a). To check its intrinsic behavior, we measured $I_{Pyro}$ with different heating rates (2, 3 and 4 K/min) and a clear rate dependence for the peak position (53.5, 56.5 and 58.5 K respectively) can be seen, unlike in ferroelectric materials [20]. Such a behavior known in the literature for other materials is usually attributed to 'thermally stimulated depolarization current (TSDC)' [21]. It is to be stressed that we are not able to detect any other peak, in particular around $T_2$ (14 K), at which there is an anomaly in the dielectric data. This establishes the destruction of ferroelectricity at $T_2$ by Sr doping. This conclusion was further verified by the bias electric method, in which we cooled the sample in the absence of an electric field, and measured the pyro-current in the warming cycle with an applied bias electric field; this protocol was recently introduced to distinguish between ferroelectricity and TSDS [22]. We are not able to detect any peak around 14 K as well as near 53.5 K (lower inset of Fig. 6a). The similar behaviors of $I_{Bias}$ and $I_{Pyro}$, confirming absence of ferroelectricity, are observed for the x= 0.2 composition as well (see figure 6(b).

Finally, we would like to make the following points: (i) The readers may see Ref. 1 for various mechanisms to explain spin-induced type-II multiferroicity in the current literature. We have also briefly discussed the same in our earlier publications [11, 15] on the parent compound. Therefore, we avoid presenting the same in detail in this publication. In a nutshell, while exchangestriction mechanism due to symmetric exchange interaction is commonly discussed for collinear spin structures [5], Dzyaloshinskii-Moriya interaction, DMI (due to asymmetric exchange interaction) is usually proposed for non-collinear spin structures such as cycloidal magnetic structural systems [23]. However, a 'local approach' [24, 25] due to canting of neighbouring spins was proposed for materials like delafossites (CuCrO$_2$) and this mechanism appears to be more general to any canted spin system – not restricted to long-range ordered magnetic structures. In Ref. 15, we have argued that Tb$_2$BaNiO$_5$ presents a new situation in which a critical canting angle (>20°) between Tb4f and Ni3d moments is required to observe canted-spin-induced ferroelectricity; as mentioned at the introduction, the magnetic sublattices of Ni and Tb are collinear and not cycloidal (or any spiral) and therefore DMI-based models are not favorable. This speaks in favor of the 'local approach' to explain Tb-Ni canted spin induced multiferroicity for the parent Tb compound. It may be mentioned that, in order to provide evidence for the existence of a critical canting angle for multiferroicity, Ram Kumar et al [26] performed neutron diffraction studies on the Sr-doped specimen (x= 0.1) and the initial studies suggest that the Tb-Ni canting angle in this specimen is less (<15°) than the critical angle value mentioned above. This observation is intriguing and provides strong evidence for the idea of critical angle, favoring the 'local approaches' as argued in Refs. 15 and 17 for multiferroicity. As emphasized by us earlier [17], on the basis of our neutron diffraction data, we are not able to resolve the question of noncentrosymmetry at present for the parent compound. (ii) It was concluded, on the basis of multiferroic transition temperature behavior due to Y substitution for Tb, that the role of Ni chain is less relevant to induce ferroelectricity, stressing the role of Tb; it is obvious from the Sr- doping studies that the role of a bigger ion like Ba in the lattice is to facilitate Tb-Ni canting beyond the critical angle. These will be elaborated by Ram Kumar et al [26] in a later publication after completion of neutron diffraction studies. As mentioned at the Introduction, the Co analogue, $Tb_2BaCoO_5$, is also a multiferroic, supporting the notion that Ni is not crucial [16] for this phenomenon. Neutron diffraction measurements are underway on this Co analogue as well to understand multiferroicity.

Summarizing, the present work reports the results of magnetization, heat capacity, dielectric, pyro-electric and magnetodielectric behavior of the Sr-doped (for Ba) specimens in $Tb_2BaNiO_5$. The key finding is that even 10 atomic percent of Sr destroys the ferroelectricity, and hence multiferroic behavior, of $Tb_2BaNiO_5$. It is also intriguing to note that there is a dramatic reduction in the maximum value of MDE in this Sr doped specimens with respect to that in the parent, though Tb sublattice is not disturbed. This result implies that cooperative electric dipole effects are also important to MDE coupling, in addition to single-ion anisotropic effects of Tb. On the basis of present studies, we believe that this compound offers an ideal opportunity for comparative doping studies at different sites by different experimental methods to get further insight of multiferroicity and magnetodielectric coupling.

One of us (EVS) would like to thank Science and Engineering Research Board, New Delhi, for awarding J C Bose Fellowship.

**References:**

1. Sang-Wook Cheong and Maxim Mostovoy, Nature materials **6,** 13 (2007).
2. T. Kimura, T. Goto, H. Shintani, K. Ishiazaka, T. Arima, and Y. Tokura, Nature (London) **426,** 55 (2003).
3. Y. Yamasaki, S. Miyasaka, Y. Kaneko, J.-P. He, T. Arima, and Y. Tokura: Phys. Rev. Lett. **96**, 207204 (2006)
4. G. Lawes et. al., Phys. Rev. Lett. **95** (2005) 087205; T. Kimura, J. C. Lashley, and A. P. Ramirez: Phys. Rev. B **73,** 220401 (R) (2006).
5. Y. J. Choi, H. T. Yi, S. Lee, Q. Huang, V. Kiryukhin, and S.-W.Cheong: Phys. Rev. Lett. **100,** 047601 (2008), and Y. Naito, K. Sato, Y. Yasui, Y. Kobayashi, Y. Kobayashi, and M. Sato: J. Phys. Soc. Jpn. **76,** 023708 (2007).
6. U. Schollwock, J. Richter, D. J. J. Farnell and R. F. Bishop, Quantum Magnetism (Springer, New York, 2004)
7. F. D. M. Haldane, Phys. Lett. **93A,** 464 (1983).
8. K. Singh, T. Basu, S. Chowki, N. Mohapatra, K K Iyer, P.L. Paulose, and E.V. Sampathkumaran, Phys Rev. B **88**, 094438 (2013).
9. T. Basu, P.L. Paulose, K.K. Iyer, K. Singh, N. Mohapatra, S. Chowki, B. Gonde, and E.V. Sampathkumaran, J. Phys.: Condens. Matter. **26,** 172202 (2014).
10. T. Basu, K. Singh, N. Mohapatra, and E.V. Sampathkumaran, J. App. Phys. **116,** 114106 (2014).
11. S. K. Upadhyay, P.L. Paulose and E.V. Sampathkumaran, Phys. Rev. B **96**, 014418 (2017).
12. J. Hernandez-Velasco and R. Saez-Puche, J. Alloys and Comp. **225,** 147 (1995).

13. E. Garcia-Matres, J.L. Garcia-Munoz, J.L. Martinez, and J. Rodriguez-Carvajal, J. Magn. Magn. Mater. **149,** 363 (1995)
14. E. Garcia-Matres, J.L. Martinez, and J. Rodriquez Carvajal, Eur. Phys. J. **B 24**, 59 (2001).
15. Ram Kumar, Sudhindra Rayaprol, Sarita Rajput, Tulika Maitra, D.T. Adroja, Kartik K Iyer, Sanjay K. Upadhyay, and E.V. Sampathkumaran, Phys. Rev. B **99,** 100406(R) (2019).
16. Sanjay K Upadhyay and E.V. Sampathkumaran, Appl. Phys. Lett. **112,** 262902 (2018).
17. Sanjay K. Upadhyay and E.V. Sampathkumaran, J. Appl. Phys. **125**, 174106 (2019).
18. J. Rodríguez-Carvajal, *Physica B* **192**, 55 (1993).
19. Vera Cuartero, Javier Blasco, J. Alberto Rodríguez-Velamazán, Joaquín García, Gloria Subías, Clemens Ritter, Jolanta Stankiewicz, and Laura Canadillas-Delgado, Phys. Rev. B **86,** 104413 (2012).
20. S.K. Upadhyay, K.K. Iyer, S. Rayaprol, P.L. Paulose, and E.V. Sampathkumaran, Sci. Rep. **6**, 31883 (2016).
21. N. Terada, Y. S. Glazkova and A. A. Belik, Phys. Rev B **93**, 155127 (2016).
22. C. De, S. Ghara, and A. Sundaresan, Solid State Commun. **205**, 61 (2015).
23. H. Katsura, N. Nagosa, and A.V. Balatsky, Phys. Rev. Lett. **95,** 057205 (2005).
24. T.A. Kaplan and S.D. Mahanti, Phys. Rev. B **83**, 174432 (2011).
25. S. Miyahara and N. Furukawa, Phys. Rev. B 93, 014445 (2016).
26. Ram Kumar, Sudhindra Rayaprol, Andreas Hoser and E.V. Sampathkumaran (to be published).

# Figure captions

**Figure.1:** X-ray diffraction pattern for **(a)** $x=0.1$, **(b)** x= 0.2, **(c)** x= 0.3, **(d)** x= 0.4 and **(e)** x= 0.5 compositions of the series $Tb_2Ba_{1-x}Sr_xNiO_5$. The continuous line through the data points are obtained by Reitveld refinement (except for x= 0.5) and difference between experimental and fitted spectra are shown in blue. Vertical bars indicate the positions for expected lines. The variation in the unit volume with Sr doping is shown in the figure **(f).** Inset of (a) shows a diffraction line to bring out that the lattice shrinks with Sr doping; an arrow is drawn to show the way the curves move with increasing x. Insets of (b), (c) and (d) show the variation of lattice parameters with the Sr doping. Asterisks in **(e)** mark extra lines.

**Figure.2: a)** Magnetic susceptibility measured with 5 kOe for $x=$ 0.1 and 0.2 compositions of the solid solution, $Tb_2Ba_{1-x}Sr_xNiO_5$, along with the data for $Tb_2BaNiO_5$, below 150 K. First derivative of these curves are shown in **(b)**. **(c)** and **(d)** show the isothermal magnetization at 5 and 30 K for x= 0.1 and 0.2 respectively, along with the corresponding curve at 5 K for the parent compound in the inset of **(d).** Inset of **(c)** and **(d)** show the ZFC-FC curves at $H=$ 100 Oe for these compositions. The dotted line in (a) is drawn to show how the feature due to $T_1$ change with x.

**Figure.3:** Heat capacity data (in the form of $C/T$) in the absence of external magnetic-field for **(a)** x= 0 and 0.1 and **(b)** x= 0.2, in the series $Tb_2Ba_{1-x}Sr_xNiO_5$. The inset of **(a)** shows the curves in an expanded form around the Neel temperature in the presence of magnetic-fields for the x= 0.1 composition and an arrow is drawn to show the direction in which the curves change with increasing field. Vertical arrows mark the two transitions.

**Figure.4:** Dielectric constant as a function of temperature, measured with different frequencies, for **(a)** x= 0.1 and **(b)** x= 0.2, in the series $Tb_2Ba_{1-x}Sr_xNiO_5$. The arrows are drawn to show the way the curves move with increasing frequency. The dielectric data measured in the presence of external magnetic fields are shown in **(b)** and **(d)** respectively. Inset of **(b)** shows the dielectric behaviour as a function of temperature for the parent compound and the curves for different frequencies (1-100 kHZ) were earlier reported to overlap.

**Figure.5:** Isothermal change in the dielectric constant with varying magnetic-field, $\Delta\varepsilon'(H)$, defined in the text, at selected temperatures for **(a)** x=0.1 and **(b)** x=0.2, in the series $Tb_2Ba_{1-x}Sr_xNiO_5$. Arrows are drawn for 5 K curve to show the virgin curve. Inset of (a) shows $\Delta\varepsilon'(H)$ virgin curve at low fields. Inset of **(b)** shows the $\Delta\varepsilon'(H)$ curve for the parent compound at $T=$ 5 K.

**Figure.6:** Pyroelectric current for **(a)** x= 0.1 and **(b)** x= 0.2 in the series $Tb_2Ba_{1-x}Sr_xNiO_5$, obtained with different heating rates as described in text. The upper insets of **(a)** and **(b)** show the pyro-current after reversing the corresponding electric field (with 2 K/min heating rate). The lower insets in **(a)** and **(b)** show the bias current data.

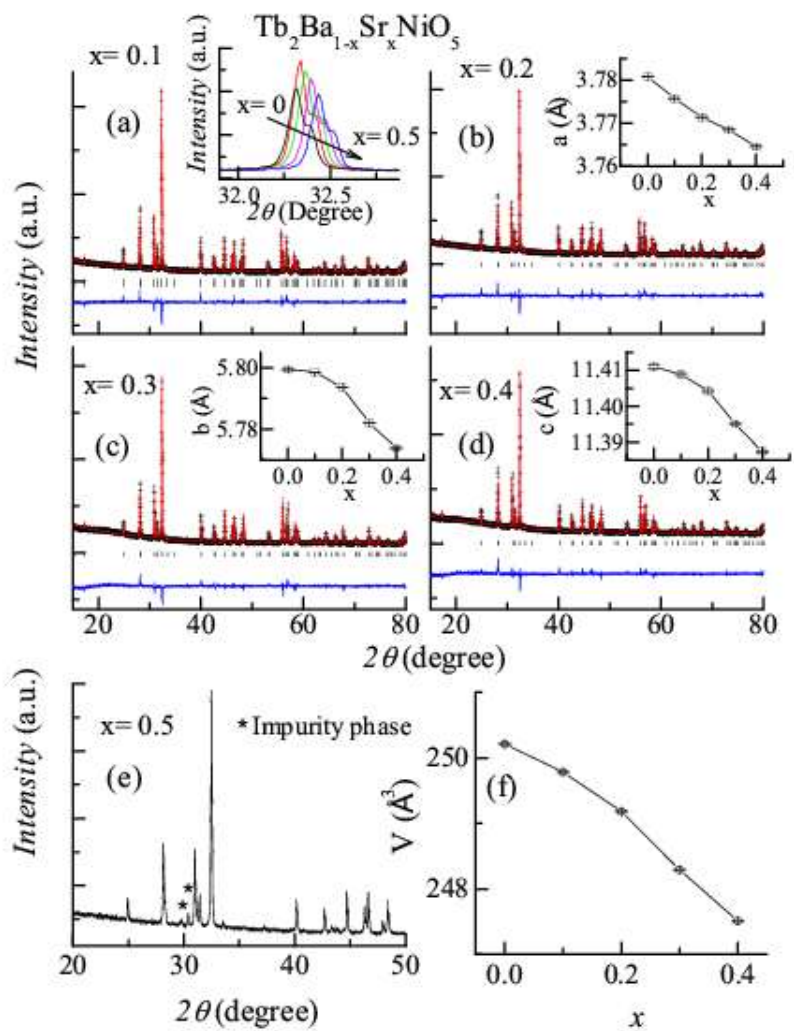

Figure 1

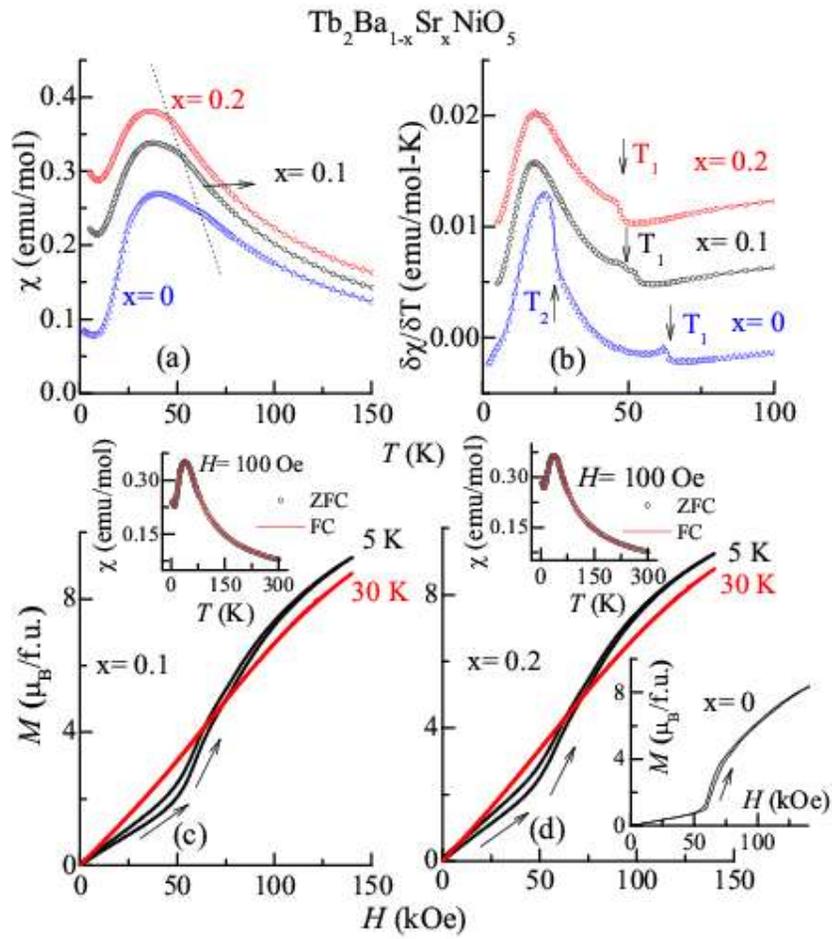

Figure 2

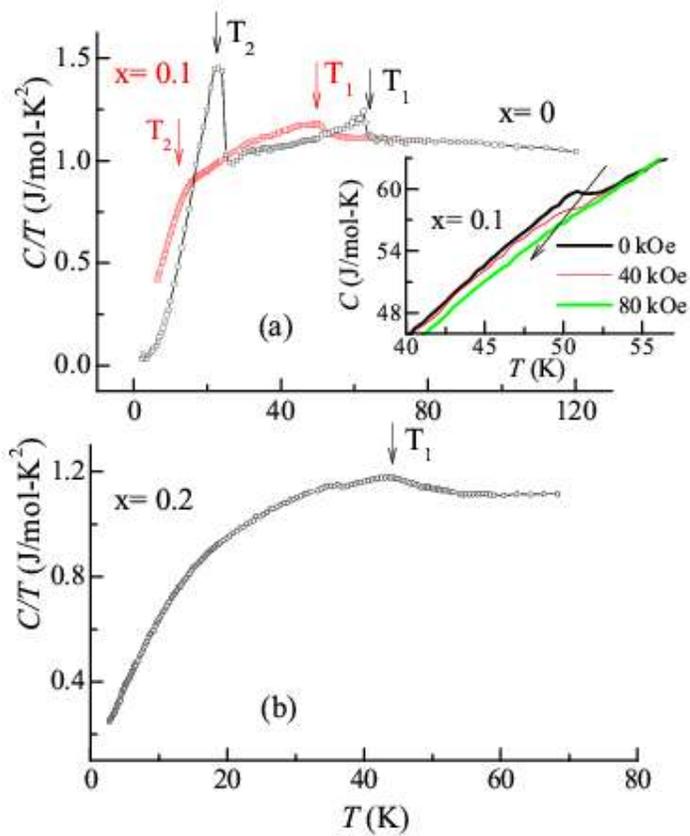

Figure 3

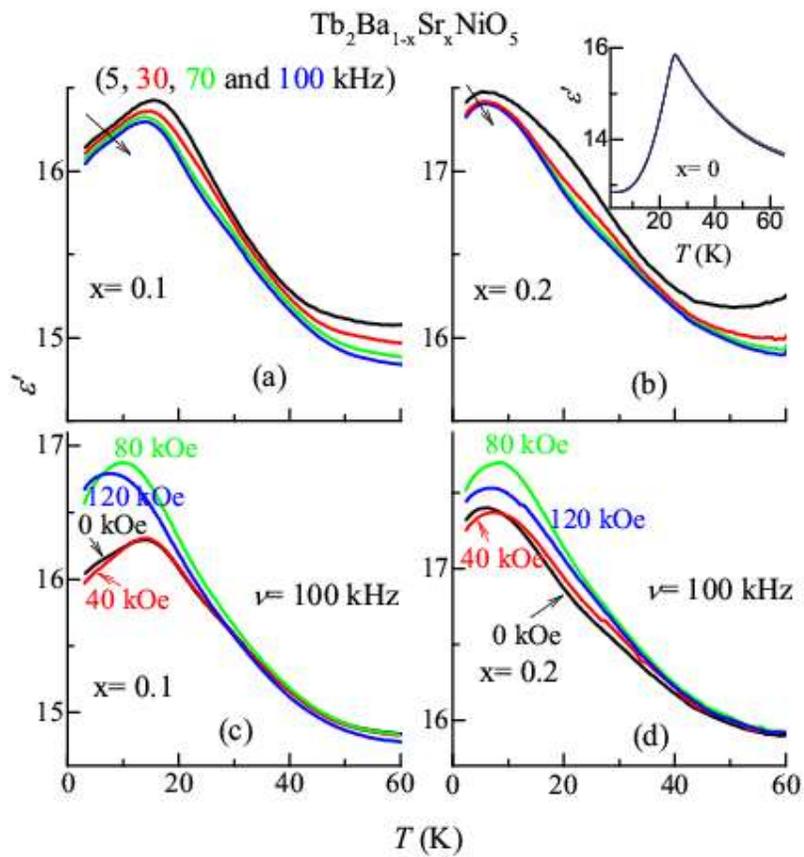

Figure 4

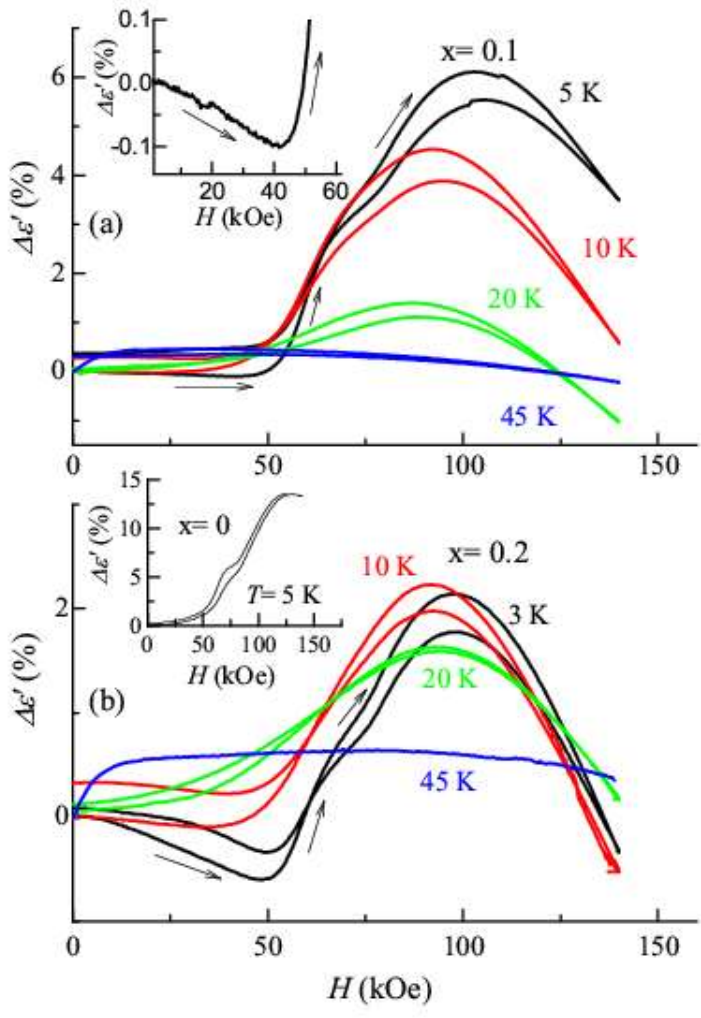

Figure 5

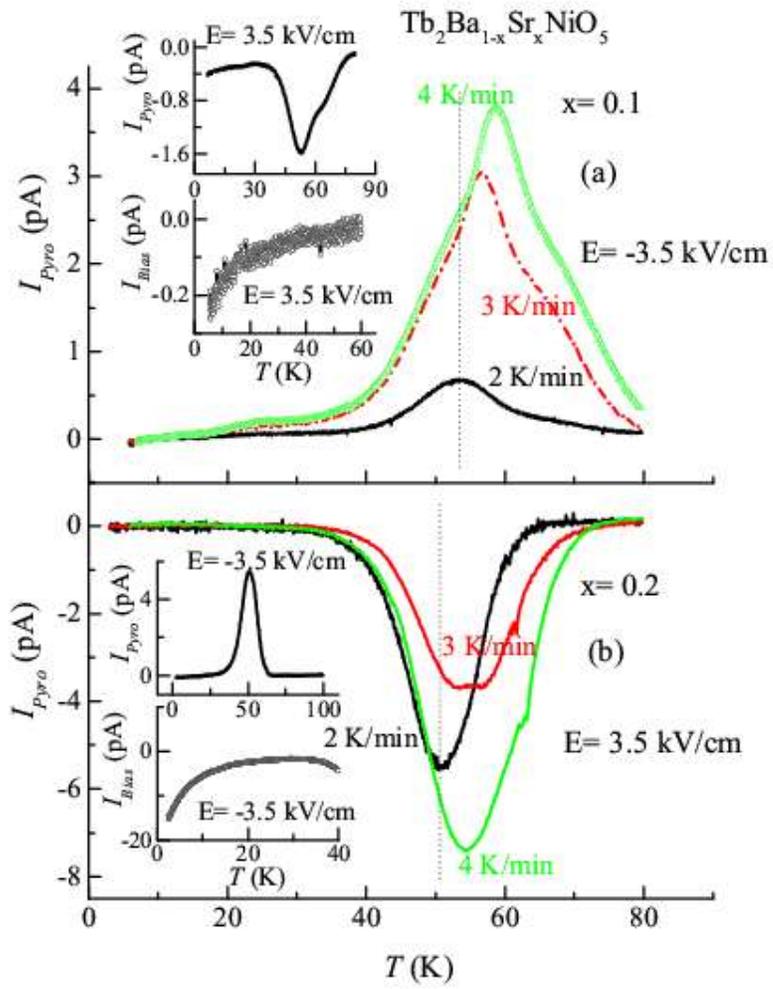